# Remark on the Comment (cond-mat/0304626) by Fal'ko, Lerner, Tsyplyatyev and Aleiner

P. Mohanty[1] and R.A. Webb[2]
[1]Department of Physics, Boston University, 590 Commonwealth Avenue, Boston, MA 02215
[2]Department of Physics, University of Maryland, College Park, MD 20742

In a recent comment[1], Fal'ko and coworkers argue that the analysis of the experimental data in ref[2] does not result in a large deviation from the prediction of One-Parameter Scaling (OPS) theory, contrary to the analysis done in ref[2]. Their arguments are based on the following claims:
(1) Extraction of cumulants in ref[2] is not done correctly.
(2) Even though Fal'ko and coworkers also find a non-zero (and relatively large) third cumulant with "their" procedure from the "raw data", the statement on the "possible violation of one-parameter scaling" in ref[2] is not warranted.
(3) The non-zero third cumulant could be explained by "the limited applicability of the ergodic theorem".

The claims, calculations and arguments in the comment by Fal'ko and coworkers are not valid because:

**(1) The authors of the comment extract the cumulants and moments from the wrong distribution.**
Fal'ko and coworkers (ref. 1) claim to have found different numbers (for moments and cumulants) from the *digitized data of histograms* shown in Fig.2 of ref.[2] (one such histogram is reproduced here as Fig.X(b)).
(a) As a matter of fact, Figs. 2 and 3 of ref.[2] display the histogram of the raw conductance data in the entire field range (-15T to +15T) at a given temperature. These histograms contain high-field, low-field and weak antilocalization contributions. In the language of random matrix theory, they have both GOE (Gaussian Orthogonal Ensemble) and GUE (Gaussian Unitary Ensemble) parts, including weak anti-localization, which contributes both to the long tail (identified in the Figs. 2 and 3) and to the central part of the peak. Clearly, values of statistical moments/cumulants extracted from the histograms in Fig.2 of ref.[2] are not the right numbers to be compared either with the values in ref.[2], or for that matter, with any theory that deals with either GOE or GUE but not both types of ensembles at the same time—the two different ensembles represent two different universality classes. A slight care in reading the legends of the Fig.2 or the text in ref.[2] would have revealed this simple fact. The cumulants/moments in ref.[2] are calculated from the raw conductance data after removing the low field part—shown, for example, as the shaded region in Fig.X(a) below, following the standard definition of cumulants/moments in ref[3] [ref.18 in the original paper in ref.[2]]. The distributions for which the moments/cumulants are reported in ref.[2] are therefore true Gaussian Unitary Ensembles.
(b) Furthermore, Fal'ko and coworkers did not even reanalyze the distribution, whether the distribution is right (GUE) or wrong (mixture). They analyzed the digitized data of a histogram. The distribution contains 9600 data points with ~800 independent sampling intervals (an independent interval is defined by the field required to apply a flux quantum h/e across a phase coherent area $L_\Phi w$). As one can clearly see in the Fig.2 of ref.[2], reproduced here as Fig.X(b), some points (bins) in the histogram could represent over 400 points from the raw distribution. Understanding that bin size in a histogram is chosen to visually best depict the nature of the distribution, the following questions become imperative. Would the cumulants extracted from a distribution (with some definition of the cumulants) be the same as the cumulants extracted from a representative histogram in which a single bin could contain over 5% of the data points? The answer is no. Would the discrepancy be more severe for higher order cumulants? The answer is yes.

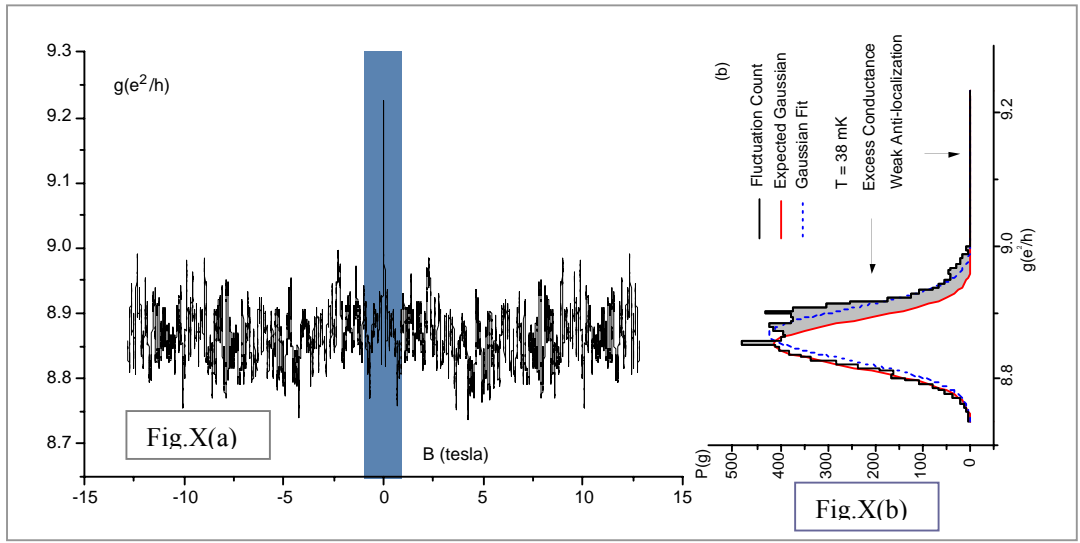

**(2) The authors of the comment ignore the phrase in the title "Possible Violation" and also the two other alternative explanations** suggested in ref[2] (page 146601-4): (a) role of interaction, and (b) validity of perturbative expansion of the scaling function β(g). To date, there is no reported analysis of a complete calculation, which includes the role of interaction on conductance distribution, and there is no reported analysis whatsoever of a nonperturbative analysis of β(g) for a disordered conductor. Furthermore, there is a large number of theoretical (and numerical) works devoted to the breakdown of one-parameter scaling. There is an equally large body of experiments showing the violation of one-parameter scaling in different types of measurements (i.e. temperature scaling in 2D systems, and metal-insulator transitions).

**(3) Limited applicability of the ergodic theorem is consistent with the values given in ref[2].**
The authors of the comment even find in their own analysis a relatively large value for the third cumulant, which—they argue, could be explained by the limited applicability of the ergodic theorem—in simple words, not enough data points for good statistics.

Variance of the cumulants $<<<g^n>>^2>$ (arising due to the limited number of sampling intervals) for n=3 is not dominated by the 3/2 power of variance. A proper calculation of the variance of the third cumulant for sample 1dD (taking into account the central limit theorem) actually results in an uncertainty due to the limited sampling length of $\sim 0.0015$. (This estimate uses N = 5 and $<<g^2>> = 1.7 \times 10^{-3}$). This uncertainty is an order of magnitude smaller than the reported uncertainty of 0.025, and the reported value of $<<g^3>>$ for 1dD, which is 0.087.

The comment on ref[2] by Fal'ko and coworkers is based on analysis on digitized data from the histograms in fig 2 and 3, and not on the distribution itself. Furthermore, the histograms depict the complete distribution with two different ensembles (GOE and GUE). Therefore, the analysis and the claims made in the comment[1] are incorrect.